\documentclass[pra, twocolumn,aps,eqsecnum,showpacs,amsmath]{revtex4}
\begin{document}
\title{ Two elementary proofs of the Wigner theorem on symmetry in 
quantum 
mechanics}
\author{R. Simon}
\email{simon@imsc.res.in}
\affiliation{The Institute of
 Mathematical Sciences, C. I. T. Campus, Chennai 600113}
\author{N. Mukunda}
\email{nmukunda@cts.iisc.ernet.in}
\affiliation{Centre for High Energy Physics, Indian Institute of 
Science, Bangalore 560012}
\author{S. Chaturvedi}
\email{scsp@uohyd.ernet.in}
\affiliation{School of Physics, University of Hyderabad, 
Hyderabad 500046}
\author{V. Srinivasan}
\email{vsspster@gmail.com}
\affiliation{Department of Theoretical Physics, University of Madras, 
Chennai 600025}
\begin{abstract}
In quantum theory, symmetry has to be defined necessarily 
in terms of the family of unit rays, the state space. The theorem 
of Wigner asserts that a symmetry so defined at the level of rays can 
always be lifted 
into a linear unitary or an antilinear antiunitary operator acting on the 
underlying Hilbert space. We present two  proofs of this theorem   
which are both elementary and economical. Central to our proofs 
is the recognition that a given  Wigner symmetry can, by 
post-multiplication by a unitary symmetry, be taken  into either the 
identity  or complex conjugation. Our analysis involves a  
 judicious interplay between the effect a given Wigner symmetry 
has on certain two-dimensional subspaces and the  effect it has on the 
entire Hilbert space.
\end{abstract} 
\maketitle 
\baselineskip12pt
\section{Introduction} 
The Wigner 
unitary-antiunitary theorem on the representation of symmetry 
operations in quantum mechanics is an important result belonging to 
the mathematical foundations of the subject.  
 The dominant role of unitary group representations in quantum physics 
 can be traced to this theorem which  
states that any 
(invertible) map of the space of pure states (unit rays) of a quantum 
system onto itself preserving transition probabilities is induced by 
either a linear unitary transformation or an antilinear antiunitary 
transformation at the level of vectors in the complex Hilbert space 
pertaining to the system. That is, invariance of all {\em transition 
 probabilities} demands that all {\em probability amplitudes} be  
either preserved or complex conjugated uniformly.

While the theorem was originally proved by 
Wigner in 1931 \cite{1}, over the decades many authors have presented 
new proofs, extensions, etc.\cite{2}-\cite{28}. Prominent among these 
is a proof by Bargmann in 1964 \cite{6} which is extremely elegant 
and, in a sense, elementary. We also mention some insightful remarks 
on this subject by Wick in 1966 \cite{7}. Some of these authors cited 
above view this 
theorem as a consequence of the fundamental theorem of projective 
geometry.

The purpose of this paper is to present two elementary proofs of the 
Wigner 
theorem which we believe have some attractive features. 
We already 
know that composition of symmetries in the Wigner sense results in 
new symmetries; and that unitary transformations on Hilbert space do 
induce Wigner symmetries on the ray space. 
 It is also clear that complex conjugation (in any chosen orthonormal 
basis) at the Hilbert space level induces a Wigner symmetry on the ray 
space. Our strategy, which is largely 
conditioned by 
our experience in classical wave optics, then in analysing a given 
Wigner symmetry is to take it to a {\em canonical form}    
by composition with a unitary symmetry {\em naturally suggested} by the 
given 
Wigner symmetry, 
and then to examine the resulting (simpler) Wigner symmetry step by step 
until it becomes completely transparent that the Wigner symmetry in its 
canonical form is  
either the identity map or complex conjugation. 

In 
this process of demonstrating that the group of all Wigner symmetries 
is the union of just two cosets with respect to the unitary group,  
as we shall see, 
there is a judicious interplay of `local' and `global' aspects, the 
former involving two-dimensional subspaces of Hilbert space and the 
latter involving general vectors not restricted to any subspace. 

The material of this paper is organised as follows.
Section II serves the dual purpose of introducing our notation  
 and making a precise statement of the problem. 
 Certain two-dimensional subspaces of the Hilbert space and their 
associated Poincar\'{e} spheres are defined. Our first proof of the Wigner 
theorem is presented in Section III in a sequence of six 
elementary steps. Some comments on the proof are presented in 
Section IV. A second proof, based on induction in the dimension $N$ of 
 the Hilbert space under consideration, is 
given in Section V,  and we conclude in Section VI with  further 
remarks. 
 
\section{Notational preliminaries, statement of the problem} 
Let ${\mathcal H}$ be the $N$-dimensional Hilbert space pertaining to 
some quantum system, with $N$ finite or infinite. Vectors and the 
inner product are denoted as usual by $|\psi\rangle,\, 
|\phi\rangle,\,\cdots,\,\langle\phi|\psi\rangle $.  The unit sphere 
${\mathcal B}$ consists of all vectors of unit length, 
\begin{equation}
 {\mathcal B}=\{~|\psi\rangle \in {\mathcal H}~\mid~\langle 
\psi|\psi\rangle=1~\}\subset {\mathcal H}. \label{2.1} 
\end{equation} 
For finite $N$, ${\mathcal B}$ is a manifold of real odd dimension  
$(2N-1)$. Rays are (equivalence) classes of vectors differing by 
phases, of the form $\{e^{i\alpha}~|\psi\rangle, 0\leq 
~\alpha~<2\pi\}$. Physical pure states correspond one--to--one to 
normalised or unit rays (i.e., to one-dimensional projections or density 
matrices). So we define 
\begin{equation}
 \mathcal{R}
=\{~\rho(|\psi\rangle)=|\psi\rangle\langle\psi|~\mid~|\psi\rangle 
\in {\mathcal B}~ \}, \label{2.2} 
\end{equation} 
and call it the ray 
space. For finite $N$, $\mathcal{R}$ has real even dimension 
$2(N-1)$. Neither ${\mathcal B}$ nor $\mathcal{R}$ is a vector 
space. There is a natural projection $\pi:{\mathcal 
B}~\rightarrow~\mathcal{R}$ given by 
\begin{equation}
 \pi\;:~\;|\psi\rangle \in {\mathcal B}~\rightarrow 
~\pi(|\psi\rangle)=\rho(|\psi\rangle)\in \mathcal{R}. \label{2.3} 
\end{equation} 
Given two rays, we have a `scalar product' 
\begin{equation} {\rm Tr}(\, \rho(|\psi_1\rangle)
 \rho(|\psi_2\rangle)\,)=|\langle\psi_1|\psi_2\rangle|^2 \geq 0,
\label{2.4}
\end{equation}
which has the standard quantum mechanical interpretation 
as transition probability.

In our first proof of Wigner's theorem, a family of two--dimensional 
subspaces of ${\mathcal H}$ plays an important role, so we define it 
now. Let $\{|n\rangle\}$, $n=1,2,\cdots,N$, be an orthonormal basis 
(ONB) for ${\mathcal H}$. With respect to this basis, for each pair 
$(j,k)$ with $j<k$ we define a two-dimensional linear subspace 
${\mathcal H}_{jk} \subset {\mathcal H}$ : 
\begin{eqnarray} 
{\mathcal H}_{jk} 
=\{~\alpha\,|j\rangle +\beta\, |k\rangle~\mid~\alpha,\beta \in 
{\mathcal C}~\}\subset {\mathcal H},&&\nonumber\\
j,k =1,2,\,\cdots,\,N,~~~j<k.&&~~
 \label{2.5} 
\end{eqnarray} 
The intersection of $\mathcal{B}$ and 
${\mathcal H}_{jk}$ is the set of unit vectors in ${\mathcal H}_{jk}$:  
 \begin{eqnarray}
 {\mathcal B}_{jk}&=&\mathcal{B}\cap {\mathcal H}_{jk}\nonumber\\ 
&=&\{\,\alpha\,|j\rangle +\beta\,|k\rangle\,\mid \,|\alpha|^2+|\beta|^2=1 
\,\}\subset {\mathcal B}. \label{2.6} 
\end{eqnarray} 
Upon projection 
this maps onto a subset $\mathcal{R}_{jk}\subset \mathcal{R}$ which 
can conveniently be parametrised using spherical polar angles on 
${\cal S}^2$: 
\begin{eqnarray} 
\mathcal{R}_{jk}&=&\pi({\mathcal 
B}_{jk}) = \{\,\rho({|\theta;\phi\rangle}_{jk})\,\}
\subset \mathcal{R},\nonumber\\ 
|\theta;\phi\rangle_{jk}&=& 
\cos\frac{\theta}{2}\,|j\rangle 
+\sin\frac{\theta}{2}~e^{i\phi}\,|k\rangle \in {\mathcal 
B}_{jk},\nonumber\\
&&~~0\leq \theta \leq 
\pi,\; 0\leq \phi < 2\pi.
\label{2.7} 
\end{eqnarray} 
Thus each ${\mathcal R }_{jk}$, $j<k$, has the form of 
a Poincar\'{e} sphere.

A {\em Wigner symmetry} (hereafter WS) is a one--to--one, onto (hence 
invertible) map $\Omega:\mathcal{R}\rightarrow \mathcal{R}$ which 
preserves the `inner product' $(\ref{2.4})$: 
\begin{eqnarray}
 \Omega&:&\rho(|\psi\rangle)\in\mathcal{R}\rightarrow 
\Omega(\,\rho(|\psi\rangle)\,)\in\mathcal{R},\nonumber\\ &&{\rm 
Tr}(\,\Omega(\,\rho(|\psi_1\rangle)\,)\Omega(\,\rho(|\psi_2\rangle)\,)\,)
=|\langle\psi_1|\psi_2\rangle|^2. 
\label{2.8} 
\end{eqnarray} 
With any such WS it is convenient to 
associate a map ${\tilde \Omega}:{\mathcal B}\rightarrow \mathcal{R}$ 
by composing $\Omega$ and $\pi$ using $(\ref{2.3})$: 
\begin{equation}
 {\tilde \Omega}=\Omega\circ \pi:{\mathcal B}\rightarrow \mathcal{R}:\;
{\tilde \Omega}(|\psi\rangle)=\Omega(\,\rho( |\psi\rangle)\,),~ 
|\psi\rangle\in{\mathcal B}. \label{2.9} 
\end{equation} 
Then the 
condition $(\ref{2.8})$ on $\Omega$ appears slightly simpler: 
\begin{equation} {\rm Tr}(\,{\tilde \Omega}(|\psi_1\rangle){\tilde 
\Omega}(|\psi_2\rangle)\,)=|\langle\psi_1|\psi_2\rangle|^2. 
\label{2.10} 
\end{equation} 
We hereafter refer to this ( or 
equivalently $(\ref{2.8})$ ) as the {\em symmetry condition} or SC.
 Clearly the composition  $\Omega_1\circ\Omega_2$ of two WS's is 
another WS.

Every unitary transformation $U$ on ${\mathcal H}$ leads to 
an associated WS 
${\,\mathcal U}$ by conjugation of density matrices: 
\begin{eqnarray} 
U:\;{\mathcal H}\rightarrow {\mathcal H}:\; \parallel 
U|\psi\rangle\parallel &=& \parallel |\psi\rangle\parallel
 \;\Rightarrow\nonumber\\ 
{\mathcal U}:\;\mathcal{R}\rightarrow 
{\mathcal R }:\; {\mathcal 
U}(\rho(|\psi\rangle))&=&\rho(U|\psi\rangle)\nonumber\\
&=&U\rho(|\psi\rangle)U^{-1}. 
\label{2.11} 
\end{eqnarray} 
Therefore given a WS $\Omega$ and a 
unitary operator $U$ on ${\mathcal H}$, by composition in either 
order we get new WS's: 
\begin{eqnarray}
 \Omega^\prime &=&{\mathcal U}\circ\Omega, ~~{\tilde \Omega}^\prime 
={\mathcal U}\circ{\tilde\Omega}; \nonumber\\ 
\Omega^{\prime\prime}&=&\Omega\circ {\mathcal U}, ~~{\tilde 
\Omega}^{\prime\prime} = \Omega\circ {\tilde {\mathcal U}}. 
\label{2.12} 
\end{eqnarray} 
We will use only the former in our 
considerations.

We can now state the problem: Given a WS $\Omega:\;\mathcal{R}
\rightarrow \mathcal{R}$, can we find or construct a one-to-one 
onto map $\omega :{\mathcal B}\rightarrow {\mathcal B}$ such that %
\begin{equation}
 {\tilde \Omega}=\Omega\circ\pi =\pi\circ\omega \label{2.13} 
\end{equation} 
corresponding to the diagram 
\begin{equation} 
\begin{array}{ccccc} & & \omega ? & & \\
                      & \mathcal{B}&\longrightarrow &\mathcal{B}& \\
                       \pi&\downarrow & &\downarrow &\pi\\
 & \mathcal{R}&\longrightarrow &\mathcal{R}& \\
 & & \Omega & & \end{array} ~~~~~~~~? \label{2.14} 
\end{equation} 
If such $\omega$ exists, we ask if it can be extended in a natural way 
from $\mathcal{B}$ to $\mathcal{H}$, and in that case how it acts on 
a general vector $|\psi\rangle\in \mathcal{H}$.

Wigner's Theorem states that this is always possible, and that $\omega$ 
extended to $\mathcal{H}$ is either a linear unitary operator $U$, so 
$\Omega=\mathcal{U}$;  or an antilinear antiunitary operator, namely 
$\omega=KU$ where $K$ is complex conjugation in some orthonormal 
basis $\{\,|n\rangle\,\}$ and $U$ is unitary. In the latter case we can 
also express $\omega$ as the product $U^\prime K$ with $U^\prime=KUK$ 
also linear unitary.

It is useful to make two remarks before we present our proofs of Wigner's 
theorem: 

Let us denote by $\{\,\mathcal{U}\,\}$ the set of 
symmetries induced by unitary transformations $U$ on the Hilbert space, 
and by  $\{\,\mathcal{UK}\,\}$ the set of  
symmetries induced by antilinear antiunitary transformations $UK$ on the 
Hilbert space.  The union 
 $\{\,\mathcal{U}\,\} \cup \{\,\mathcal{UK}\,\}$   thus constitutes 
a set of obvious symmetries in the sense of Wigner. Wigner's theorem is 
the assertion that {\em  there 
are no more symmetries beyond these obvious ones}.

Maps which are positive but not completely 
positive are fundamental to  
quantum information theory.  
Our second remark is concerning the implication of Wigner's theorem 
in this context: $\{\,\mathcal{U}\,\}$ is the set of {\em all} 
completely positive maps which map the family of pure states onto itself.    
Since complex conjugation  $\{\,\mathcal{K}\,\}$  and matrix 
transposition $\{\,\mathcal{T}\,\}$  are equivalent maps at the level of 
density operators we have:  
 $\{\,\mathcal{UT}\,\}$ 
is the set of {\em all} 
positive but not completely positive maps which map the family of pure 
states onto itself. That is, {\em any other positive but not completely 
positive map will take some pure states into mixed states}.    

\section{First Proof of Wigner's theorem }

Let a WS $\Omega$ with its associated map 
$\tilde{\Omega}:\mathcal{B}\rightarrow\mathcal{R}$ be given. It is 
clear from eqs. $(\ref{2.8})$, $(\ref{2.9})$ that for any 
$|\psi\rangle \in \mathcal{B}$ we have 
\begin{equation}
 \tilde{\Omega}(|\psi\rangle)\equiv\Omega(\pi(|\psi\rangle))
=\pi(|\psi^\prime\rangle), 
\label{3.1} 
\end{equation} 
where $|\psi^\prime\rangle\in\mathcal{B}$ 
is determined upto a phase. We use this fact repeatedly in the 
following. The proof is made up of six steps, each quite elementary. 
We now present them in sequence. 

\vskip0.2cm 
\noindent 
{\em Step 1}\,:~ Choose some (any) ONB $\{|n\rangle\}$ for 
$\mathcal{H}$. From the SC $(\ref{2.10})$ for pairs of basis vectors and 
eq. $(\ref{3.1})$ it follows that 
\begin{equation} 
\tilde{\Omega}(|n\rangle)=\pi(|n;\Omega\rangle),~ n=1,2,\,\cdots,\,N, 
\label{3.2} \end{equation} 
where each $|n;\Omega\rangle$ is 
determined upto a phase and the collection $\{|n;\Omega\rangle\}$ 
also is an ONB for $\mathcal{H}$. Make some (any) choices for the 
vectors $|n;\Omega\rangle$ and define a unitary transformation $U$ on 
$\mathcal{H}$ by 
\begin{equation} U\,|n;\Omega\rangle =|n\rangle,~ 
n=1,2,\cdots,N. \label{3.3} 
\end{equation} 
We now define a WS 
$\Omega^\prime$ by 
\begin{equation}
 \Omega^\prime=\mathcal{U}\circ\Omega,~~ 
~\tilde{\Omega}^\prime=\mathcal{U}\circ\tilde{\Omega}, \label{3.4} 
\end{equation} 
and reduce the analysis of $\Omega$ to that of 
$\Omega^\prime$. This WS has a simple action on the vectors  
$\{\,|n\rangle\,\}$, namely, 
\begin{equation} 
\tilde{\Omega}^\prime(|n\rangle)=\pi(|n\rangle),~ n=1,2,\,\cdots,\,N, 
\label{3.5} 
\end{equation} 
and by the SC $(\ref{2.10})$ for $|n\rangle$ 
and general $|\psi\rangle\in\mathcal{B}$ we have 
\begin{equation} 
\tilde{\Omega}^\prime(\sum_{n=1}^N c_n\,|n\rangle)=\pi(\sum_{n=1}^N 
c_n^\prime\,|n\rangle), |c_n^\prime|=|c_n|. \label{3.6} 
\end{equation} 
To repeat, 
{\em the symmetry} $\tilde{\Omega}^\prime$
 {\em leaves invariant the standard set of orthonormal rays} 
$\{\,\pi(|n\rangle)\,\}$, {\em and this can be arranged for any Wigner 
symmetry} $\Omega$.  

We hasten to add that 
 $\Omega^\prime=\mathcal{U}\circ\Omega$  is not yet the intended 
canonical form for $\Omega$. The reason for this is the fact that  
the unitary operator $U$ could have been  post-multiplied  by 
{\em any}  
diagonal unitary operator, still leaving invariant the standard set of 
orthonormal rays. The canonical form for $\Omega$ will indeed get fixed 
once we exercise (and exhaust) this freedom in Step~3.

\vskip0.2cm 
\noindent 
{\em Step 2}\,:~ Next we limit ourselves to 
vectors in $\mathcal{B}_{jk}$ for a chosen  $j<k$, and study the 
action 
of $\tilde{\Omega}^\prime$ on such vectors. It is in fact adequate to 
look at the action on a (latitude) circle of vectors 
$|\theta_0;\phi\rangle_{jk}$, eq. $(\ref{2.7})$, for fixed 
$\theta_0\in(0,\pi)$ and varying $\phi \in [0,2\pi)$. From eq.  
$(\ref{3.6})$ we see that 
\begin{eqnarray} 
\tilde{\Omega}^\prime(|\theta_0;\phi\rangle_{jk})
=\pi(|\theta_0;\phi^\prime\rangle_{jk}), 
\label{3.7} 
\end{eqnarray} 
with $\theta_0,j,k$ unchanged and 
$\phi^\prime$ dependent on $\phi$ ( and possibly also on 
$\theta_0,j,k$) in an invertible manner. Since 
\begin{eqnarray}
|_{jk}\langle\theta_0;\phi_1|\theta_0;\phi_2\rangle_{jk}|^2
 &=& 
2\cos^2\frac{\theta_0}{2}\sin^2\frac{\theta_0}{2}\cos(\phi_1-\phi_2)\nonumber\\
&&~~+\cos^4\frac{\theta_0}{2}+\sin^4\frac{\theta_0}{2}\,,
\label{3.8} 
\end{eqnarray} 
use of the SC $(\ref{2.10})$ for this pair 
of vectors on the latitude circle $\theta_0$ shows that $\phi_1,\phi_2$ 
are 
carried 
by 
$(\ref{3.7})$ into $\phi_1^\prime,\phi_2^\prime$ such that 
\begin{equation} 
\cos(\phi_1^\prime-\phi_2^\prime)=\cos(\phi_1-\phi_2). \label{3.9} 
\end{equation} 
It follows that the change $\phi\rightarrow\phi^\prime 
$ in eq. $(\ref{3.7})$ caused by $\tilde{\Omega}^\prime$ action is of 
the form 
\begin{eqnarray} \phi^\prime&=&\phi_{jk}+\epsilon_{jk}\phi, 
~\phi_{jk}\in [0,2\pi),~ \epsilon=\pm 1. \label{3.10} 
\end{eqnarray} 
(This and similar later equations are understood to be valid 
mod~$2\pi$.) Thus the action $(\ref{3.7})$ reads 
\begin{equation} 
\tilde{\Omega}^\prime(|\theta_0;\phi\rangle_{jk})
=\pi(|\theta_0;\phi_{jk}+\epsilon_{jk}\phi\rangle_{jk}). 
\label{3.11} 
\end{equation} 
For fixed $\theta_0,j,k$ this means that 
$\tilde{\Omega}^\prime$ acts on $\phi$ via some element of the group 
$O(2)$ : $\phi_{jk}$ denotes an $SO(2)$ element, while 
$\epsilon_{jk}$ determines whether we have a proper or improper 
rotation. 

\vskip0.2cm 
\noindent 
{\em Step 3}\,:~ We now use the freedom, noted at the end of Step~1,  
to multiply each vector $|n\rangle$ in the ONB 
$\{|n\rangle\}$ by an independent phase factor, preserving the 
structure of the results obtained upto this point. Thus we define a 
(diagonal) unitary transformation $U^\prime$ on $\mathcal{H}$ by 
\begin{equation}
 U^\prime \,|n\rangle=e^{-i\phi_n}\,|n\rangle,~n=1,2,\,\cdots,\,N, 
\label{3.12} 
\end{equation} 
and pass from the WS $\Omega^\prime$ to 
the (final) WS $\Omega^{\prime\prime}$ by 
\begin{equation} 
\Omega^{\prime\prime}=\mathcal{U}^\prime\circ\Omega^\prime,~~~ 
\tilde{\Omega}^{\prime\prime}=\mathcal{U}^\prime\circ\tilde{\Omega}^\prime.
 \label{3.13} 
\end{equation} 
This helps us simplify the phases or  
$SO(2)$ angles $\phi_{jk}$ in eq. $(\ref{3.11})$ to some extent by 
suitable choices of the $\phi_n$. We have : 
\begin{eqnarray}
 \tilde{\Omega}^{\prime\prime}(|\theta_0;\phi\rangle_{jk})
&=&\tilde{\mathcal{U}}^\prime(|\theta_0; 
\phi_{jk}+\epsilon_{jk}\phi\rangle_{jk})\nonumber\\ 
&=&\pi(|\theta_0;\phi_{jk}+\phi_j-\phi_k+\epsilon_{jk}\phi\rangle_{jk}).~~~\nonumber\\ 
\label{3.14} 
\end{eqnarray} 
Remembering that $k\geq2$, we choose
 \begin{equation} \phi_1=0,~\phi_n=\phi_{1n}~~~{\rm for}~n\geq 2. 
\label{3.15} 
\end{equation} 
Then we have 
\begin{eqnarray} 
\tilde{\Omega}^{\prime\prime}(|\theta_0;\phi\rangle_{jk})&=& 
\pi(|\theta_0;\phi_{jk}^\prime+\epsilon_{jk}\phi\rangle_{jk}),\nonumber\\ 
\phi_{jk}^\prime &=&\phi_{jk}+\phi_j-\phi_k,\nonumber\\ 
\phi_{1k}^\prime &=&0,~~k\geq 2. \label{3.16} 
\end{eqnarray} 
Compared 
to eq. $(\ref{3.11})$, the sign factors $\epsilon_{jk}$ are unchanged, 
while the $SO(2)$ angles $\phi_{jk}$ have been simplified to 
$\phi_{jk}^\prime$ with $\phi_{1k}^\prime=0$.

\vskip0.2cm 
\noindent 
{\em Step 4}\,:~ Since 
$\Omega^{\prime\prime}$ is related to $\Omega^{\prime}$ by the 
diagonal unitary transformation $U^\prime$, eq. $(\ref{3.11})$, it is 
clear that the structure of eq. $(\ref{3.6})$ is retained for 
$\tilde{\Omega}^{\prime\prime}$: 
\begin{equation} 
\tilde{\Omega}^{\prime\prime}(\,\sum_{n=1}^N 
c_n\,|n\rangle\,)=\pi(\,\sum_{n=1}^N c_n^{\prime\prime}\,|n\rangle\,),~ 
~|c_n^{\prime\prime}|=|c_n|. \label{3.17} 
\end{equation} 
Now we show 
that in the action $(\ref{3.16})$ by $\tilde{\Omega}^{\prime\prime}$, 
not only is  $\phi_{1k}^\prime =0$ but in fact $\phi_{jk}^\prime =0$ for 
all $j<k$. This is a very important consequence of the SC 
$(\ref{2.10})$. 
We choose a single special `real' (normalized) vector 
$\mbox{\boldmath{$r$}}^{(0)} =
(\,r_1^{(0)},\,r_2^{(0)},\,r_3^{((0)},\,\cdots\,\,)^T$,  
and define $|\mbox{\boldmath{$r$}}^{(0)}\rangle\in 
\mathcal{B}$ with the following form:  
\begin{equation}
|\mbox{\boldmath{$r$}}^{(0)}\rangle =\sum_{n=1}^{N}r_n^{(0)}\,|n\rangle, ~~ 
 r_n^{(0)} ~{\rm real ~and}\; \ne \,0.
\label{3.18} 
\end{equation}
Under $\tilde{\Omega}^{\prime\prime}$ action 
we have, from eq. $(\ref{3.17})$, 

\begin{equation}
 \tilde{\Omega}^{\prime\prime}(\,|{\mbox{\boldmath{$r$}}^{(0)}\rangle}\,)
=\pi(\,\sum_{n=1}^{N}r_n^{(0)}e^{i\eta_n}\,|n\rangle\,), 
\label{3.19} 
\end{equation} 
for some phases $\eta_n$. We now invoke 
the SC $(\ref{2.10})$ for the pair of vectors $
|\theta_0;\phi\rangle_{jk},\; |\mbox{\boldmath{$r$}}^{(0)}\rangle$, any 
$j<k$,  
under $\tilde{\Omega}^{\prime\prime}$ action to get :
\begin{eqnarray*}  
|r_j^{(0}e^{i\eta_j}\cos\frac{\theta_0}{2} + r_k 
e^{i\eta_k}\sin\frac{\theta_0}{2}~e^{-i(\phi_{jk}^\prime
+\epsilon_{jk}\phi)}|^2&&
\nonumber\\ 
=|r_j^{(0)}\cos\frac{\theta_0}{2}
+r_k^{(0)}\sin\frac{\theta_0}{2}~e^{-i\phi}|^2\,.&&\nonumber\\ 
\end{eqnarray*}  
That is,   
\begin{eqnarray}  
\cos(\epsilon_{jk}\phi+\phi_{jk}^\prime+\eta_j-\eta_k)=\cos\phi,~ 
0\leq\phi<2\pi.~ \label{3.20} 
\end{eqnarray} 
This implies 
\begin{equation} 
\phi_{jk}^\prime=\eta_k-\eta_j,~j<k. \label{3.21} 
\end{equation} 
For $j=1$, from eq.$(\ref{3.16})$ we have 
$\eta_k=\eta_1$ independent of $k$. Putting this back into eq. 
$(\ref{3.21})$ gives $\phi_{jk}^\prime=0$ for all $j<k$. Thus the 
actions $(\ref{3.16})$, $(\ref{3.20})$ simplify to 
\begin{eqnarray} 
\tilde{\Omega}^{\prime\prime}(\,|\theta_0;\phi\rangle_{jk}\,)&=& 
\pi(\,|\theta_0;\epsilon_{jk}\phi\rangle_{jk}\,), \nonumber\\ 
\tilde{\Omega}^{\prime\prime}(\,|\mbox{\boldmath{$r$}}^{(0)}\rangle\,)
&=&\pi(\,|\mbox{\boldmath{$r$}}^{(0)}\rangle\,). 
\label{3.22} 
\end{eqnarray} 
For each pair $j<k$, only a sign factor 
$\epsilon_{jk}$ remains. We will soon prove that this 
factor cannot depend on $j,\,k$.

 \vskip0.2cm 
\noindent 
{\em Step 5}\,:~ Let 
us next invoke the SC $(\ref{2.10})$ under $\Omega^{\prime\prime}$ 
for the pair of vectors $|{\mbox{\boldmath{$c$}}}\rangle$, 
$|\theta_0;\phi\rangle_{jk}$ where $|{\mbox{\boldmath{$c$}}}\rangle$ is 
the 
general normalised linear combination occurring on the left hand sides  
of eqs. $(\ref{3.6})$, $(\ref{3.17})$. On the basis of the results 
$(\ref{3.17})$, $(\ref{3.22})$ we have: 
\begin{eqnarray} 
|c_j^{\prime\prime}\cos\frac{\theta_0}{2}
+c_k^{\prime\prime}\sin\frac{\theta_0}{2}~e^{-i\epsilon_{jk}\phi}|^2 
~~~~~~~~~~~~&&\nonumber\\
=|c_j\cos\frac{\theta_0}{2}+c_k 
\sin\frac{\theta_0}{2}~e^{-i\phi}|^2,&&\nonumber\\
{\rm i.e.},~ c_j^{\prime\prime}c_k^{\prime\prime 
*}e^{i\epsilon_{jk}\phi}+c_j^{\prime\prime *}c_k^{\prime\prime 
}e^{-i\epsilon_{jk}\phi} ~~~~~~~~~~~&&\nonumber\\
= c_j c_k^{*}e^{i\phi}+c_j^{ *}c_k 
e^{-i\phi},
 ~~\phi\in [0,2\pi).&&~~ \label{3.23} 
\end{eqnarray} 
Therefore 
the transition $\{c_n\}\rightarrow \{c_n^{\prime\prime}\}$ must 
follow 
\begin{eqnarray} 
c_j^{\prime\prime}c_k^{\prime\prime*}
&=& c_j c_k^{*}~{\rm 
if}~\epsilon_{jk}=+1,\nonumber\\ 
c_j^{\prime\prime}c_k^{\prime\prime*}
&=&c_j^{*} c_k~ {\rm if}~\epsilon_{jk}=-1. \label{3.24} 
\end{eqnarray} 
\vskip0.2cm 
\noindent 
{\em Step 6}\,:~ This is the 
final step in the proof. We show that consistency demands that the choice 
of 
$\epsilon_{jk}$ cannot depend on the pair $j,\,k$\,; it must be 
uniformly 
$+1$ or uniformly $-1$ for all pairs.
 Choose any $j,k,\ell$ with $j<k<\ell$. For any vector 
$|\mbox{\boldmath{$c$}}\rangle \in \mathcal{B}$, we have the elementary 
result 
\begin{equation} c_j c_k^{*}c_k c_l^{*} (c_j 
c_\ell^{*})^*=|c_j c_k c_\ell|^2= {\rm real}\geq 0 . \label{3.25} 
\end{equation} 
Therefore if $|\mbox{\boldmath{$c$}}\rangle$ is taken by 
$\tilde{\Omega}^{\prime\prime}$ action to 
$|\mbox{\boldmath{$c$}}^{\prime\prime}\rangle$ as in eq.  $(\ref{3.17})$ 
we 
must necessarily have 
\begin{equation}
 c_j^{\prime\prime}c_k^{\prime\prime *}\cdot 
c_k^{\prime\prime}c_\ell^{\prime\prime *} \cdot 
(c_j^{\prime\prime}c_\ell^{\prime\prime *})^*\,=\,{\rm real}\,\geq\, 0. 
\label{3.26} 
\end{equation} 
Depending on the values of 
$\epsilon_{jk}, \epsilon_{k\ell}, \epsilon_{j\ell}$, by eq. 
$(\ref{3.24})$ this requires that, whatever $\{c_n\}$ may be 
\begin{equation} (c_j c_k^{*} ~{\rm or }~c_j ^{*}c_k)\cdot (c_k 
c_\ell^{*} {\rm or }~c_k ^{*}c_\ell) \cdot (c_j c_l^{*} {\rm or 
}~c_j ^{*}c_\ell)^* = {\rm real}\geq 0. \label{3.27} 
\end{equation} 
In each factor we have the first expression for $\epsilon=1$, the 
second for $\epsilon=-1$. It is now immediate that if 
$\epsilon_{jk}=\epsilon_{k\ell}=\epsilon_{j\ell}=+1$, or if 
$\epsilon_{jk}=\epsilon_{k\ell}=\epsilon_{j\ell}=-1$, this condition 
is obeyed. But in every other case, the left hand side of eq. 
$(\ref{3.27})$ is an expression which is in general complex. 
Therefore we have the final result:
\begin{eqnarray} 
\tilde{\Omega}^{\prime\prime}(\,\sum_{n=1}^{N}c_n\,|n\rangle\,)
&=&\pi(\,\sum_{n=1}^{N}c_n\,|n\rangle\,)~{\rm 
if~ all}~\epsilon_{jk}=+1,\nonumber\\ 
&=&\pi(\,\sum_{n=1}^{N}c_n^*\,|n\rangle\,)~{\rm 
if~ all}~\epsilon_{jk}=-1.~~~~~~ \label{3.28} 
\end{eqnarray} 
These are the only consistent possibilities. For action on vectors in 
$\mathcal{B}$, the originally given WS $\Omega$ is thus either the 
product $\mathcal{U}^{-1}\circ\mathcal{U}^{\prime-1}$ of unitary WS's 
or the product 
$\mathcal{U}^{-1}\circ\mathcal{U}^{\prime-1}\circ\mathcal{K}$, where 
$\mathcal{K}$ is the WS corresponding to complex conjugation in 
the ONB $\{|n\rangle\}$. These immediately extend from $\mathcal{B}$ 
to $\mathcal{H}$ in the natural ways, thus establishing Wigner's 
Theorem.

\section{Extensions of WS actions on special vectors}

In the course of the proof of Wigner's theorem given in the preceding 
Section, important roles were played by the specially chosen vectors 
$|\theta_0;\phi\rangle_{jk}\in\mathcal{B}_{jk}\subset\mathcal{H}_{jk}$ in 
each two-dimensional subspace defined by eqs. $(\ref{2.5})$,  
$(\ref{2.6})$; and the `real' vector $|\mbox{\boldmath{$r$}}^{(0)}\rangle$ 
of 
eq. $(\ref{3.18})$. We emphasize that only one (latitude) circle of 
vectors in $\mathcal{B}_{jk}$, and a single `real' vector $ 
|\mbox{\boldmath{$r$}}^{(0)}\rangle$, were actually used in the proof. 
However it is easy to show that, at corresponding steps of the proof, 
Steps 2 and 4, we can obtain more information about the actions of 
$\tilde{\Omega}^{\prime},\,\tilde{\Omega}^{\prime\prime}$ respectively.

Consider two values $\theta_0,\theta_0^{\prime}$ of the polar angle 
over ${\cal S}^2$ and let eq. $(\ref{3.11})$ in the two cases read 
\begin{eqnarray}
 \tilde{\Omega}^\prime(\,|\theta_0;\phi\rangle_{jk}\,)
&=&\pi(\,|\theta_0;\phi_{jk}+\epsilon_{jk}\phi 
\rangle_{jk}\,),\nonumber\\ 
\tilde{\Omega}^\prime(\,|\theta_0^\prime;
\phi^\prime\rangle_{jk}\,)&=&\pi(\,|\theta_0^\prime;
\phi_{jk}^\prime+\epsilon_{jk}^\prime\phi^\prime\rangle_{jk}\,). 
\label{4.1} 
\end{eqnarray} 
The SC $(\ref{2.10})$ for the two vectors on 
the left here is : 
\begin{eqnarray*} 
|\cos\frac{\theta_0}{2}\cos\frac{\theta_0^\prime}{2} + 
\sin\frac{\theta_0}{2}\sin\frac{\theta_0^\prime}{2}~ 
e^{i(\epsilon_{jk}\phi+\phi_{jk}-\epsilon_{jk}^\prime 
\phi^\prime-\phi_{jk}^\prime)}|^2 &&\nonumber\\ 
= |\cos\frac{\theta_0}{2}\cos\frac{\theta_0^\prime}{2} + 
\sin\frac{\theta_0}{2}\sin\frac{\theta_0^\prime}{2}~ 
e^{i(\phi-\phi^\prime)}|^2\,.&&\nonumber\\ 
\end{eqnarray*}
That is, 
\begin{eqnarray} 
{\rm i.e.},\; \cos(\phi-\phi^\prime)=
\cos(\epsilon_{jk}\phi-\epsilon_{jk}^\prime\phi^\prime
+\phi_{jk}-\phi_{jk}^\prime),&&\nonumber\\
~0\leq\phi,\,\phi^\prime<2\pi.~~~&& 
\label{4.2} 
\end{eqnarray} 
This immediately implies 
\begin{equation} 
\epsilon_{jk}^\prime=\epsilon_{jk},~\phi_{jk}^\prime=\phi_{jk},
 \label{4.3} 
\end{equation} 
so in fact we have in place of eq.  
$(\ref{3.11})$: 
\begin{eqnarray}
 \tilde{\Omega}^\prime(|\theta;\phi\rangle_{jk})
&=&\pi(|\theta;\phi_{jk}+\epsilon_{jk}\phi\rangle_{jk}),\nonumber\\ 
&&~~~~0\leq\theta\leq\pi,~~0\leq\phi <2\pi, \label{4.4} 
\end{eqnarray} 
with 
$\phi_{jk},~\epsilon_{jk}$ constant over ${\cal S}^2$.

Next, in connection with the result at Step 4, we easily see that we 
have the wider result 
\begin{eqnarray}
|\mbox{\boldmath{$r$}}\rangle&=&\sum_{n=1}^{N}r_n\,|n\rangle\in \mathcal{B},~    
r_n ~{\rm real, ~and}\; r_1 \ne0\nonumber\\ 
&&\Rightarrow\,
\tilde{\Omega}^{\prime\prime}(|\mbox{\boldmath{$r$}}\rangle)
=\pi(|\mbox{\boldmath{$r$}}\rangle) 
\label{4.5} 
\end{eqnarray} 
for all `real' vectors of this form.

\section{Second Proof of Wigner's Theorem}
The crucial  step in the proof of the Wigner theorem given in Section 
III is the  first one -- the passage from $\Omega$ to $\Omega^\prime$ in 
eq. $(\ref{3.4})$ , resulting in eqs. $(\ref{3.5})$, 
$(\ref{3.6})$. That this is so is clearly brought out  by the second 
proof we now present, based on induction in the dimension of the Hilbert 
space.  

For a two-dimensional quantum system, as is so well known and used 
earlier, the  space $\mathcal{R}$ is the Poincar\'e sphere ${\cal S}^2$. 
As in 
eq.  $(\ref{2.7})$,  each pure state density matrix corresponds 
one--to--one to a point on this sphere, and we have:
\begin{eqnarray}
&& \hat{\mbox{\boldmath{$n$}}}\in {\cal S}^2\rightarrow 
\rho(\hat{\mbox{\boldmath{$n$}}})=
\frac{1}{2}(1+\hat{\mbox{\boldmath{$n$}}} \cdot\mbox{\boldmath 
{$\sigma$}}),\nonumber\\
&&{\rm Tr}(\rho(\hat{\mbox{\boldmath{$n$}}}_1)  
\rho(\hat{\mbox{\boldmath{$n$}}}_2))=
\frac{1}{2}(1+\hat{\mbox{\boldmath{$n$}}}_1\cdot  
\hat{\mbox{\boldmath{$n$}}}_2).
\label{5.1}
\end{eqnarray}
 (Orthogonal states correspond to antipodal points). Thus a WS in this 
case is a one--to--one onto map 
$\Omega\,:~{\cal S}^2\rightarrow {\cal S}^2$ preserving angles between 
pairs of 
points:
\begin{equation}
\Omega(\hat{\mbox{\boldmath{$n$}}}_1)\cdot  
\Omega(\hat{\mbox{\boldmath{$n$}}}_2)=\hat{\mbox{\boldmath{$n$}}}_1 
\cdot \hat{\mbox{\boldmath{$n$}}}_2.
\label{5.2}
\end{equation}
As is very well known, such maps are either proper rotations belonging 
 to $SO(3)$, induced by two dimensional unitary transformations of 
$U(2)$ on the underlying two--dimensional 
 Hilbert space $\mathcal{H}^{(2)}$; or they are improper rotations in 
 $O(3)$, involving in addition complex conjugation on 
$\mathcal{H}^{(2)}$ (mirror reflection on ${\cal S}^2)$. This is Wigner's 
theorem in this case. 

 For general finite dimension we use induction. We assume the theorem is 
 true in $N$ dimensions, then prove it for $(N+1)$ dimensions. Let 
 ${\mathcal H}^{(N+1)}$ be an $(N+1)$ dimensional Hilbert space, and 
 $\Omega$ a WS for the corresponding quantum system. Choose any ONB 
 $\{\,|j\rangle, ~j=1,\,2,\,\cdots,\,N+1\,\}$ for $\mathcal{H}^{(N+1)}$. 
As 
in 
the 
 initial steps of Section III, pass from $\Omega$ to another WS 
 $\Omega^\prime$ by composition with a suitable unitary symmetry. Then 
as in eqs. $(\ref{3.5})$, $(\ref{3.6})$ we have:
\begin{eqnarray}
&&\tilde{\Omega}^\prime(\,|j\rangle\,) 
=\pi(\,|j\rangle\,),~~~j=1,\,2,\,\cdots,\,N+1\,;\nonumber\\
&& \tilde{\Omega}^\prime(\,\sum_{j=1}^{N+1}c_j|j\rangle\,) 
=\pi(\,\sum_{j=1}^{N+1}c_j^\prime|j\rangle\,),~~~|c_j|=|c_j^\prime|. 
\label{5.3}
\end{eqnarray}
 Let us identify the $N$-dimensional subspace $\mathcal{H}^{(N)}\subset 
 \mathcal{H}^{(N+1)}$ as the subspace spanned by the first $N$ basis 
 vectors:
\begin{equation}
{\cal H}^{(N)}={\rm Sp}\{\,|j\rangle, j=1,2,\cdots,N\,\}\subset 
 {\cal H}^{(N+1)}.
\label{5.4}
\end{equation}
By eq.$(\ref{5.3})$, $\Omega^\prime$ is a WS for $\mathcal{H}^{(N)}$. 
By our assumption, may be after composition with a diagonal unitary 
symmetry which we suppress for simplicity, we have either 
\begin{eqnarray*}
\tilde{\Omega}^\prime(\,\sum_{j=1}^{N}c_j |j\rangle\,)
 =\pi(\,\sum_{j=1}^{N}c_j|j
\rangle\,),~~\forall\;\mbox{\boldmath{$c$}}\equiv(c_1,\cdots, 
c_N)^T\,, 
\end{eqnarray*}
or
\begin{eqnarray}
\tilde{\Omega}^\prime(\,\sum_{j=1}^{N}c_j |j\rangle\,)
 =\pi(\,\sum_{j=1}^{N}c_j^{*}|j
\rangle\,),~~\forall\;\mbox{\boldmath{$c$}}\,. 
\label{5.5}
\end{eqnarray}

 Suppose the first  choice holds. To deal with $\mathcal{H}^{(N+1)}$, a 
general vector here is
\begin{eqnarray}
&& |\mbox{\boldmath{$c$}}\rangle + c_{N+1}|N+1\rangle\in 
\mathcal{H}^{(N+1)},\nonumber\\
&&|\mbox{\boldmath{$c$}}\rangle =\sum_{j=1}^{N}c_j|j\rangle \in 
\mathcal{H}^{(N)}.
\label{5.6}
\end{eqnarray}
Let
\begin{eqnarray}
 \tilde{\Omega}^\prime(\,|\mbox{\boldmath{$c$}}\rangle + 
 c_{N+1}|N+1\rangle\,)&=&\pi(\,|\mbox{\boldmath{$c$}}^\prime\rangle + 
c_{N+1}^\prime|N+1\rangle\,),\nonumber\\
|c_j^\prime|=|c_j|,&&j=1,2,\cdots,N+1.
\label{5.7}
\end{eqnarray}
The SC $(\ref{2.10})$ for a general $|\mbox{\boldmath{$c$}}^{(1)}\rangle 
\in \mathcal{H}^{(N)}$ and the vector $(\ref{5.6})$ in  
$\mathcal{H}^{(N+1)}$ yields
\begin{eqnarray}
&&|\,\sum_{j=1}^{N} c_j^{(1)*}c_j^\prime\,|^2 =  |\,\sum_{j=1}^{N} 
c_j^{(1)*}c_j\,|^2, ~\;\forall\;\mbox{\boldmath{$c$}}^{(1)}
\nonumber\\
&&~~\Rightarrow\mbox{\boldmath{$c$}}^{(1)\dagger} 
\mbox{\boldmath{$c$}}^\prime 
\mbox{\boldmath{$c$}}^{\prime\dagger}\mbox{\boldmath{$c$}}^{(1)} =
\mbox{\boldmath{$c$}}^{(1)\dagger} \mbox{\boldmath{$c$}} 
\mbox{\boldmath{$c$}}^{\dagger}\mbox{\boldmath{$c$}}^{(1)},~\; 
\forall\;\mbox{\boldmath{$c$}}^{(1)} \nonumber\\
&&~~~~~~\Rightarrow \mbox{\boldmath{$c$}}^\prime 
= ({\rm phase})\;\mbox{\boldmath{$c$}}.
\label{5.8}
\end{eqnarray}
Then eq. $(\ref{5.7})$ simplifies to 
\begin{eqnarray}
\tilde{\Omega}^\prime(|\mbox{\boldmath{$c$}}\rangle +  
c_{N+1}|N+1\rangle)&=&\pi(|\mbox{\boldmath{$c$}}\rangle +  
c_{N+1}^{\prime\prime}|N+1\rangle),\nonumber\\
&&|c_{N+1}^{\prime\prime}|=|c_{N+1}|.
\label{5.9}
\end{eqnarray}
Next the SC for two vectors in $\mathcal{H}^{(N+1)}$ yields:
\begin{eqnarray*}
|\mbox{\boldmath{$c$}}^{(1)}\rangle + c_{N+1}^{(1)}|N+1\rangle , 
~~|\mbox{\boldmath{$c$}}^{(2)}\rangle + c_{N+1}^{(2)}|N+1\rangle
\in \mathcal{H}^{(N+1)},&&\nonumber\\
|\,\mbox{\boldmath{$c$}}^{(1)\dagger}\mbox{\boldmath{$c$}}^{(2)}
+c_{N+1}^{(1)\prime\prime*}c_{N+1}^{(2)\prime\prime}\,|^2
=|\,\mbox{\boldmath{$c$}}^{(1)\dagger}\mbox{\boldmath{$c$}}^{(2)}+
c_{N+1}^{(1)*}c_{N+1}^{(2)}\,|^2,&&\nonumber\\
 \forall\;
\mbox{\boldmath{$c$}}^{(1)},
\mbox{\boldmath{$c$}}^{(2)}.
\end{eqnarray*}
That is,
\begin{eqnarray}
c_{N+1}^{(1)\prime\prime*}c_{N+1}^{(2)\prime\prime}
  = c_{N+1}^{(1)*}c_{N+1}^{(2)}.
\label{5.10}
\end{eqnarray}
So
\begin{eqnarray}
c_{N+1}^{\prime\prime}&=&e^{i\phi}c_{N+1},\nonumber\\
\tilde{\Omega}^\prime(\,|\mbox{\boldmath{$c$}}\rangle + 
c_{N+1}|N+1\rangle\,)&=&\pi(\,|\mbox{\boldmath{$c$}}\rangle 
+ e^{i\phi}c_{N+1}|N+1\rangle\,),\nonumber\\
\label{5.11}
\end{eqnarray}
the phase $\phi$ being independent of $|\mbox{\boldmath{$c$}}\rangle$. 
Then by a diagonal unitary phase transformation we can pass to a WS 
 $\tilde{\Omega}^{\prime\prime}$ which acts trivially on 
 $\mathcal{H}^{(N+1)}$. This proves, in this case, that validity of 
Wigner's theorem  
in $N$ dimensions implies its validity in $(N+1)$ dimensions.

For the second option  in $(\ref{5.5})$, a similar 
 argument holds.  Validity of Winger's theorem for 
$N=2$ is evident as noted in the opening paragraphs of this Section. 
 Hence, by induction, proof of Wigner' theorem is complete for all 
 $N = 2,\,3,\,\cdots\,$.
  
We conclude this Section with some additional  remarks on the 
infinite-dimensional case.  
For a WS $\Omega$ on a Hilbert space  $\mathcal{H}$ of infinite 
dimension, choose an ONB  $\{\,|n\rangle,~n=1,\,2,\,\cdots\,\, \}$,
denote by 
$\mbox{\boldmath{$\psi$}}$ a column vector 
$(\,\psi_1,\,\psi_2,\,\psi_3,\,\cdots\,\,)^T$, 
 and pass to $\Omega^\prime$ such that
\begin{eqnarray}
\tilde{\Omega}^\prime(\,|n\rangle\,)
 =\pi(\,|n\rangle\,),~n&=&1,\,2,\,\cdots,\;\nonumber\\
\tilde{\Omega}^\prime(\,|\psi\rangle
 =\sum_{n=1}^{\infty}\psi_n |n\rangle\,)&=&\pi(\,|\psi^\prime\rangle
=\sum_{n=1}^{\infty}\psi_n ^\prime|n\rangle\,); 
\nonumber\\
&&~~~~|\psi_n^\prime|=|\psi_n|.
\label{5.12}
\end{eqnarray}
This ONB gives a sequence of subspaces 
$\mathcal{H}^{(2)}\subset \mathcal{H}^{(3)}\,\cdots\, 
\subset  \mathcal{H}^{(N)}\subset \mathcal{H}^{(N+1)}\,\cdots\,$. 
So as in the above we can pass from the WS $\Omega^\prime$ to 
an $\Omega^{\prime\prime}$  whose action on $\mathcal{H}^{(N)}$ 
is either trivial for all finite $N$ or is complex conjugation 
for all finite $N$. Now the SC for 
$|\mbox{\boldmath{$c$}}\rangle\in\mathcal{H}^{(N)},\; 
|\mbox{\boldmath{$\psi$}}\rangle \in\mathcal{H}$ in the trivial case is 
\begin{eqnarray}
\tilde{\Omega}^{\prime\prime}(\,|\mbox{\boldmath{$c$}}\rangle 
\;{\rm or}\;  |\mbox{\boldmath{$\psi$}}\rangle\,) &=&
\pi(\,|\mbox{\boldmath{$c$}}\rangle 
\;{\rm or}\;  |\mbox{\boldmath{$\psi$}}^\prime\rangle\,),\nonumber\\ 
&&~~|\psi_n^\prime|=|\psi_n|,\;\,\forall\, n\,;\nonumber\\
\mbox{\boldmath{$c$}}^\dagger\mbox{\boldmath{$\psi$}}^\prime
\mbox{\boldmath{$\psi$}}^{\prime\dagger}\mbox{\boldmath{$c$}}&=&
\mbox{\boldmath{$c$}}^\dagger\mbox{\boldmath{$\psi$}}
\mbox{\boldmath{$\psi$}}^{\dagger}\mbox{\boldmath{$c$}}\,.
\label{5.13}
\end{eqnarray}                            
While $\mbox{\boldmath{$\psi$}}\mbox{\boldmath{$\psi$}}^\dagger$ is 
in general infinite dimensional, $(\ref{5.13})$ holds for all 
$\mbox{\boldmath{$c$}}$ for all finite $N$, hence 
\begin{equation}
\mbox{\boldmath{$\psi$}}^\prime = ({\rm phase}) 
\,\mbox{\boldmath{$\psi$}},
\end{equation}
and the triviality of $\tilde{\Omega}^{\prime\prime}$ 
action is established on $\mathcal{H}$. A similar argument 
holds when complex conjugation on all $\mathcal{H}^{(N)}$ 
is needed.

\section{Concluding remarks} 
We hope to have provided  new elementary proofs of 
 Wigner's unitary-antiunitary theorem on the representation of symmetry 
operations in 
quantum mechanics which are both elementary and economical on the one 
hand, and effectively combine `local' and `global' aspects on the 
other. By the latter we mean that while for the most part we deal 
with the action of a Wigner Symmetry on vectors in certain 
two-dimensional subspaces of the full Hilbert space ${\mathcal H}$, 
at all stages there is a clear and evolving understanding of the 
action on general vectors in ${\mathcal H}$. This may well be 
contrasted with the well known and extremely elegant proof by 
Bargmann; there, at every stage, the arguments work with just two or 
three  vectors, so to that extent, the global picture seems missing.

In connection with Steps 3 and 4 of the proof in Section III, the 
following comment may be made. At the start of Step 3, there is a 
`matrix' of phases $\phi_{jk}$ still to be examined; and the diagonal 
unitary transformation $U^\prime$ involves a much smaller number of 
phases $\phi_n$. Therefore it is understandable that by Step 3, which 
does not use the SC $(\ref{2.10})$, it is only possible to transform 
the phases $\phi_{1k}$ to zero. Step 4, however, by use of the SC 
$(\ref{2.10})$ is able to show that after the action of $U^\prime$, 
{\em all} the phases $\phi_{jk}$ have been transformed to zero. This 
seems 
to be a particular feature of the present first proof, bringing out the 
power of the SC $(\ref{2.10})$ in a direct fashion.

The second proof of Section V is different in spirit from 
 the first proof, though both rest on the idea of 
passing from $\Omega$ to $\Omega^\prime$. 
This is reminiscent of `passing to the rest frame' or to 
a `local inertial frame' in relativistic problems. 
After this first step, however, the two proofs are structured 
differently, though both are quite elementary. 

Finally, it is clear that the set of all Wigner symmetries of a system 
of Hilbert 
space dimension $N$ forms a group. This group is not twice as large as 
$U(N)$ or $SU(N)$. The centre $Z_N$ of $SU(N)$ has $N$ elements,  
all of which are multiples of identity; and these elements leave every 
ray invariant. Thus the relevant group is not $SU(N)$, but the quotient 
 $SU(N)/Z_N$; and the adjoint representation of 
$SU(N)$ is indeed a faithful representation of this quotient group. 
[\,It 
is again in view of this nontrivial centre that 
$U(N)$ is not 
$SU(N)\times U(1)$ but the quotient $(\,SU(N)\times U(1)\,)/Z_N$.\,]   
 It  is well known, and already noted at the beginning of Section V,  
that in the case $N=2$  the group of Wigner 
symmetries is the union of two copies of $SO(3)$. The point being made 
is that this feature is true for all $N$, with $SO(3) = SU(2)/Z_2$ 
replaced by $SU(N)/Z_N$.

\noindent
{\em Acknowledgement}: The authors would like to thank P.P. Divakaran 
for many insightful discussions and, in particular, for drawing the 
authors' attention to the work of Wick\cite{7}. 
V. Srinivasan would like to acknowledge
an Emeritus Fellowship of the University Grants Commission
which made this work possible.


\begin{thebibliography}{99} %\section*{References and footnotes} 
\bibitem{1} E. P. Wigner, {\it Gruppentheorie}, {Vieweg, 
Braunschweig, 1931} pp 251-254; {\it Group Theory}, {Academic Press 
Inc., New York, 1959}, pp 233-236. 

\bibitem{2} R. Hagedorn, ``A note on symmetry operations in quantum 
mechanics'',  Nuovo 
Cimento {\bf XII}, 553-566 (1959). 

\bibitem{3} U. Uhlhorn,  ``Representation of symmetry 
transformations in quantum mechanics'', Arkiv der 
Physik {\bf 23}, 307-340(1962). 

\bibitem{4} J.S. Lomont and P. 
Mendelson, ``The Wigner unitary-antiunitary theorem'', Ann. Math. {\bf 
78}, 548-559 (1963). 

\bibitem{5} G. Emch 
and C. Piron, ``Symmetry in quantum theory'',  J. Math. Phys. {\bf 4}, 
469-473(1963). 

\bibitem{6} V. 
Bargmann, ``Note on Wigner's theorem in quantum mechanics'', J. Math. 
Phys. {\bf 5}, 862-868(1964). 

\bibitem{7} G. C. 
Wick, ``On symmetry transformations'', in {\it Preludes in Theoretical 
Physics}, in honour of V.F. 
Weisskopf, edited by   A . de Shalit, H. 
Feshbach, and L. van Hove, {North  Holland, Amsterdam, 1966}. 

\bibitem{8} V. S. Varadarajan, {\it Geometry of Quantum Theory}, (van 
Nostrand, 1968), Vol.~I, Chapter~VII, Sec.~3.  

\bibitem{9} L. Bracci, G. Morchio, F. Strocchi, 
``Wigner's theorem on symmetries in indefinite metric spaces'', 
Commun. Math. Phys. {\bf 41}, 289-299 (1975). 

\bibitem{10}
 B. Simon, ``Quantum dynamics: from automorphism to hamiltonian'', in  
{\em Studies in 
Mathematical
Physics: Essays in Honour of Valentine Bargmann}, edited by 
E.H. Lieb, B. Simon, and A.S. Wightman
  (Princeton University Press, 1976), pp. 327--349.

\bibitem{11}
R. Wright, ``The structure of projection-valued states: A generalization 
of Wigner's theorem'', Int. J. Theor. Phys. {\bf 16}, 567--573 (1977).

\bibitem{12}
L.C. Biedenharn and J.D. Louck, {\em The Racah-Wigner Algebra in Quantum 
Theory}, Vol.~9 of  {\em Encyclopedia in Mathematics and its 
Applications}, (Addison-Wesley, 1981), Chap.~5.

\bibitem{13}
P.M. Van den Broek, ``Symmetry transformations in indefinite metric 
spaces: A generalization of Wigner's theorem'', Physica~A {\bf 127}, 
599--612 (1984). 

\bibitem{14}
N. Gisin, ``Generalization of Wigner's theorem for dissipative quantum 
systems'', J. Phys. A: Math. Gen. {\bf 19}, 205--210 (1986).


\bibitem{15}
C.S. Sharma and D.F. Almeida, ``A direct proof of Wigner's theorem on 
maps which preserve transition probabilities between pure states of 
quantum mechanics'', Ann. Phys. {\bf 197}, 300--309 (1990). 

\bibitem{16} S. Weinberg, 
{\it The Theory of Quantum Fields} Vol.~I ( CUP, 1995), 
 Appendix A, pp.91--96. 

\bibitem{17}
L. Moln\'{a}r, ``Wigner's unitary-antiunitary theorem via Herstein's 
theorem
on Jordan homeomorphisms'', J. Nat. Geom. {\bf 10}, 137--148 (1996).

\bibitem{18}
J. R\"{a}tz, ``On Wigner's theorem: Remarks, complements, comments, and 
corollaries'', Aequationes Math. {\bf 52}, 1--9 (1996).

\bibitem{19} 
G. Cassinelli, E. DeVito, P. J. Lahti, and A. Levrero, 
 ``Symmetry groups in quantum mechanics and the theorem of Wigner 
on the symmetry transformations'', 
Rev. Math. Phys. {\bf 4}, 921-941 (1997). 

\bibitem{20}
L. Moln\'{a}r, ``An algebraic approach to Wigner's unitary-antiunitary 
theorem'', J. Aus. Math. Soc. A. Pure Math. Stat. {\bf 65}, 354--369 
(1998); \\
 ``A generalization of Wigner's unitary-antiunitary 
theorem to Hilbert modules'', Jour. Math. Phys.. 
{\bf  40}, 5544--5554 (1999); \\ 
 ``Generalization of Wigner's unitary-antiunitary 
theorem for indefinite inner-product spaces'', Commun. Math. Phys.. 
{\bf  210}, 785--791 (2000);\\ 
 ``Orthogonality preserving transformations on indefinite 
inner product spaces: generalization of Uhlhorn's version of Wigner's 
theorem'',   
 J. Funct. Anal. {\bf  194}, 248--262 (2002). 


\bibitem{21}
C.-A. Faure, ``An elementary proof of the fundamental theorem of 
projective geometry'', Geom. Dedicata {\bf 90}, 145--151 (2002).    

\bibitem{22}
 D. Bakic and B. Guljas, ``Wigner's theorem in  Hilbert 
$C^*$-modules over $C^*$-algebras of compact operators'', Proc. Am.  
Math. Soc.  {\bf 130}, 2343--2349 (2002);\\
 ``Wigner's theorem in a class of Hilbert 
$C^*$-modules'', Jour. Math. Phys. {\bf 44}, 2186--2191 (2003).

\bibitem{23}
P. Semrl, ``Generalized symmetry transformations on quaternionic 
indefinite inner product spaces: An extension of quaternionic 
version of Wigner's theorem'', Commun. Math. Phys. {\bf 242}, 
579--584 (2003).

\bibitem{24}
M. Gyory, ``A new proof of Wigner's theorem'', 
Rep.  Math. Phys. 
 {\bf 54}, 159--167 (2004).
 
\bibitem{25}
G. Chevalier, ``Lattice approach to Wigner-type theorems'',  
Int. J. Theor. Phys. {\bf 44}, 1905--1915 (2005). 

\bibitem{26}
G. Chevalier, ``Wigner-type theorems for projections'',  
Int. J. Theor. Phys. {\bf 47}, 69--80 (2008); 

\bibitem{27} K.J. Keller, N.A.  
Papadopoulos and A.F. Reyes-Lega,``On the realizations of symmetries in 
quantum mechanics'', Math. Semesterber. {\em online version}: DOI 
10.1007/s00591-08-0035-5.  arXiv:0712.099 [quant-ph].

\bibitem{28}
M. Buth, ``A simple proof of the theorem of Wigner'', 
arXiv:0802.3624 [math-ph]. 
\end{thebibliography}
\end{document}